\documentclass[apj]{emulateapj}

\usepackage{graphicx}
\usepackage{amssymb}
\usepackage{amsmath}
\usepackage{natbib}
\usepackage{xcolor}
\def\beq{\begin{equation}} \def\eeq{\end{equation}}
\def\bea{\begin{eqnarray}} \def\eea{\end{eqnarray}}

\def\nn{\nonumber}
\def\L{{\rm L}}\def\U{{\rm U}}

  \def\mir{\mathrm{r}} \def\mit{\mathrm{\theta}}

\graphicspath{{pic/}}
\def\ricpic{0.33\hsize}

\begin{document}

\title{Controversy of the GRO J1655-40 black hole mass and spin estimates and its possible solutions}

\author{Z.~Stuchl\'{\i}k$^{1}$\thanks{zdenek.stuchlik@fpf.slu.cz} and M.~Kolo\v{s}$^{1}$\thanks{martin.kolos@fpf.slu.cz}}

\affiliation{
$^{1}$Institute of Physics and Research Centre of Theoretical Physics and Astrophysics, Faculty of Philosophy \& Science, Silesian University in Opava,\\ Bezru\v{c}ovo n\'{a}m\v{e}st\'{i} 13, CZ-74601 Opava, Czech Republic}

\begin{abstract}
Estimates of the black hole mass $M$ and dimensionless spin $a$ in the microquasar GRO J1655-40 implied by strong gravity effects related to the timing and spectral measurements are controversial, if the mass restriction determined by the dynamics related to independent optical measurements, $M_{\rm opt}=(5.4\pm0.3) M_{\odot}$, are applied. The timing measurements of twin high-frequency (HF) quasiperiodic oscillations (QPOs) with frequency ratio $3:2$ and the simultaneously observed low-frequency (LF) QPO imply the spin in the range $a\in(0.27-0.29)$ if models based on the frequencies of the geodesic epicyclic motion are used to fit the timing measurements, and correlated creation of the twin HF QPOs and the LF QPO at a common radius is assumed. On the other hand, the spectral continuum method implies $a\in(0.65-0.75)$, and the Fe-line-profile method implies $a\in(0.94-0.98)$. This controversy can be cured, if we abandon the assumption of the occurrence of the twin HF QPOs and the simultaneously observed LF QPO at a common radius. We demonstrate that the epicyclic resonance model of the twin HF QPOs is able to predict the spin in agreement with the Fe-profile method, but no model based on the geodesic epicyclic frequencies can be in agreement with the spectral continuum method. We also show that the non-geodesic string loop oscillation model of twin HF QPOs predicts spin $a>0.3$ under the optical measurement limit on the black hole mass, in agreement with both the spectral continuum and Fe-profile methods.
\end{abstract}


\maketitle

\section{Introduction}\label{intro}
Strong gravity affecting an accretion disc in vicinity of a black hole horizon governs three observationally significant phenomena enabling to determine the black hole mass $M$ and the dimensionless spin $a$. The phenomena are related to the timing effects, namely to the frequencies of the HF QPOs and connected LF QPOs, to the spectral continuum of the accretion disc, and the Fe spectral lines profiled by the influence of the black hole spacetime. 

The most interesting (and precise) information is connected to the twin HF QPOs observed with the fixed frequency ratio $3:2$ in the microquasars GRS 1915+105, XTE J1550-56, and GRO J1655-40. Such twin HF QPOs can be explained by the so called geodesic oscillation models using frequencies of the geodesic epicyclic motion in the field of Kerr black holes, i.e., the orbital frequency, $\nu_{\phi}$, and the epicyclic radial, $\nu_r$, and latitudinal, $\nu_{\theta}$, frequencies. However, the twin HF QPOs in the three microquasars cannot be explained by a fixed oscillation model, if we assume a Kerr black hole \citep{Tor-etal:2011:ASTRA:}. A unique, epicyclic resonance model exists \citep{Kot-etal:2014:ASTRA:} if we assume central Kerr naked singularities that demonstrate special properties of the prograde circular motion \citep{Stu:1980:BAC:,Stu-Sche:2012:CLAQG:}.

In the case of the GRS 1915+105 microquasar, the $3:2$ HF QPOs can be explained by the epicyclic resonance model \citep{Tor-etal:2005:ASTRA:} \footnote{To explain whole the frequency set of the HF QPOs observed in the GRS 1915+105 microquasar, more complex models have to be invoked \citep{Stu-Sla-Tor:2007:ASTRA:,Stu-etal:2005:PHYSR4:}}, while for the XTE J1550-56 and GRO J1655-40 microquasars, the twin HF QPOs and the related LF QPO can be explained by the relativistic precession model \citep{Ste-Vie-Mor:1999:ApJ:} that is by definition combined with the relativistic nodal model of the LF QPOs \citep{Ste-Vie:1998:ApJ:}. 

In the case of the microquasar GRS 1915+105 the limits on the black hole mass and spin implied by the models of the twin HF QPOs are in agreement with the limits implied by the spectral measurements \citep{McCli-Rem:2004:CompactX-Sources:}, and agreements of limits implied by models of the QPOs and the models of spectral measurements has been demonstrated also in the case of the XTE J1550-56 microquasar \citep{Mot-etal:2014b:MNRAS:}. On the other hand, in the case of the microquasar GRO J1655-40, the spin limits of the spectral (continuum and Fe-line) measurements \citep{Sha-etal:2006:ApJ:,Mil-etal:2009:ApJ:} are contradicting each other, and, moreover, they are both contradicting the spin limits implied by the geodesic models of QPOs \citep{Mot-etal:2014a:MNRAS:,Stu-Kol:2016:ASTRA:}, if the limits on the black hole mass determined by the dynamical studies based on optical measurements \citep{Bee-Pod:2002:MNRAS:} are considered. 

In the present paper we consider the controversial restrictions on the GRO J1655-40 black hole mass and spin. Keeping the relevance of the mass restrictions implied by the weak gravity dynamical models based on the optical measurements of the binary system that have high relevance and are quite independent of the timing and spectral measurements connected with the strong gravity near the black hole horizon, we first discuss the possibility to obtain an agreement of the twin HF QPO geodesic models and any of the spectral methods, and then we test possibility of agreement of the string loop oscillation model of twin HF QPOs with the predictions of the spectral measurements. 

\section{The controversy of the mass and spin limits on the GRO J1655-40 black hole} 

The GRO J1655-40 low-mass-X-ray-binary (LMXB) source is one of the extensively studied Galactic microquasars, i.e., the sources where the accreting central object is assumed to be a black hole. However, controversial estimates of the GRO J1655-40 black hole gravitational mass $M$ and dimensionless spin $a$ have been reported recently \citep{Mot-etal:2014a:MNRAS:,Stu-Kol:2016:ASTRA:}. We first summarize origin of this controversy related in the strong gravity regime to models of X-ray spectral measurements implying limits on the black hole spin, and the so called geodesic models of QPOs implying in the strong gravity regime limits on the black hole mass and spin that are confronted with the limits on the black hole mass implied by dynamical studies in the weak gravity regime. 

Mass of the GRO J1655-40 black hole is restricted by dynamical methods related to spectro-photometric optical measurements \citep{Bee-Pod:2002:MNRAS:} that are not related to the timing studies of QPOs based on the X-ray measurements, and have high degree of credibility as they are related to the weak field gravity techniques. The range of allowed values of the black hole mass implied by the optical measurements reads \citep{Bee-Pod:2002:MNRAS:}  
\beq
     M_{\rm opt} = (5.4\pm0.3)~M_{\odot}. \label{Mopt}
\eeq

Note that there exists an earlier and larger estimate of the GRO J1655-40 black hole mass determined in \cite{Oro-Bai:1997:ApJ:} that reads 
\beq
     M_{\rm bh} = (7.02\pm0.22)~M_{\odot}. \label{Mopt2}
\eeq
However, the later estimate presented in \cite{Bee-Pod:2002:MNRAS:} is in recent papers considered as the relevant one.

The X-ray spectroscopy techniques give in the case of the GRO J1655-40 microquasar controversial results. The spectral continuum measurements \citep{Sha-etal:2006:ApJ:} predict the black hole spin in the range 
\beq
     0.65 < a < 0.75 . \label{a-cont}
\eeq
The measurements of the Fe spectral lines profiled by the strong gravity of the black hole predict \citep{Mil-etal:2009:ApJ:} 
\beq
     0.94 < a < 0.98 . \label{a-line}
\eeq
Clearly, there is a strong discrepancy between the results of spectral measurements related to radiation processes in the strong gravity regime. We could expect that the timing effects related to QPOs occuring in the strong gravity regime could help to determine the correct spectral measurements due to the spin estimates. 

The Rossi XTE observatory brings many timing measurements of the X-rays emitted by the GRO J1655-40 source that are summarized in \cite{Mot-etal:2014b:MNRAS:}. Interesting are those related to the twin HF QPOs and the LF QPOs. The LF QPOs were observed between 0.1~Hz and 30~Hz, usually independently on the HF QPOs. However, most important is observation of twin HF QPOs with frequencies stabilized at the lower frequency $\nu_{\L}\sim300$~Hz and the upper frequency $\nu_{\U}\sim450$~Hz that were observed simultaneously; then it is natural to assume that the twin HF QPOs are related to a common radius where both the observed oscillatory modes occur. The twin HF QPOs were reported for the first time in \cite{Str:2001:ApJ:} where simultaneously with the twin HF QPOs a LF QPO was reported at $\nu_{\rm LF}\sim17$~Hz. The magnitude of the observed frequencies and the character of the oscillations indicate strongly that the twin HF QPOs should occur in close vicinity of the black hole horizon, being related to the orbital motion. 

The special simultaneous observation of the twin HF QPOs and the LF QPO enables to obtain stringent restrictions on the mass and dimensionless spin of the central black hole, if we assume that all of these QPOs arise at a given radius of the accretion disc \citep{Mot-etal:2014a:MNRAS:}. The Monte Carlo technique applied to the observational data and the relativistic precession (RP) model, belonging to the geodesic models of twin HF QPOs, with the frequency identification $\nu_{\L}=\nu_{\phi}-\nu_{r}$ and $\nu_{\U}=\nu_{\phi}$, along with the related relativistic nodal precession model of the LF QPO with the frequency identification $\nu_{\rm nod}=\nu_{\phi}-\nu_{\theta}$, implied for the black hole mass and spin the limits \citep{Mot-etal:2014a:MNRAS:} 
\beq
     M = (5.31\pm0.07)~M_{\odot},\quad a = 0.290\pm0.003. \label{M-a-mot}
\eeq

The mass limit is in very good agreement with the mass limit given by the optical measurements. However, there is a clear discrepancy with the spin limits given by both the spectral measurements, as the RP model predicts spin $a<0.3$. 

Recently, the standard nodal precession model of the LF QPOs and a variety of the geodesic models of twin HF QPOs, i.e., models using frequencies of the oscillatory modes combined from the frequencies of the geodesic epicyclic motion, has been tested for matching to observational data of the simultaneously observed twin HF QPOs and LF QPO in the microquasar GRO J1655-40; models originally proposed to explain only the twin HF QPOs were generalized to interpret also the LF QPO by the relativistic nodal precession \citep{Stu-Kol:2016:ASTRA:}. Instead of the Monte Carlo technique, the method of frequency relations introduced in \cite{Stu-Kot-Tor:2013:ASTRA:} has been used \citep{Stu-Kol:2015:MNRAS:}, and the fitting has been done to the data of the twin HF QPOs and simultaneous LF QPO presented as Sample B1 in Tab.2. of \cite{Mot-etal:2014a:MNRAS:}. The peak frequencies with measurement error of the peak frequencies (centroid frequencies dominated by statistics of the frequencies
 ) were used \citep{Mot-etal:2014a:MNRAS:} 
\bea
 &&\nu_{\rm L} = 298\pm4~\mathrm{Hz}, \quad \nu_{\rm low} = 17.3\pm0.1~\mathrm{Hz}, \nonumber\\
 &&\nu_{\rm U} = 441\pm2~\mathrm{Hz}. \label{fff}
\eea

For all the considered models of twin HF QPOs it was assumed that both the twin HF QPOs and the LF QPO arise at a common radius. Then the frequency relation technique enables to obtain a mass-spin relation $M_{\rm HF}(a,p)$ for the twin HF QPOs with frequency ratio parameter 
\beq 
     p = \left(\frac{\nu_\L}{\nu_\U}\right)^2 ,
\eeq 
and due to the assumption of the common radius a mass-spin relation $M_{\rm LF}(a,p)$ can be determined for the LF QPO. The two relations imply the limits on the black hole mass and spin related to the QPO measurements \citep{Stu-Kol:2016:ASTRA:}. It has been found in \cite{Stu-Kol:2016:ASTRA:} that three geodesic models can predict the mass in agreement with the optical measurement limit. Along with the RP models when the frequency relation method implies the limits 
\beq
     M = (5.3\pm0.1)~M_{\odot},\quad a = 0.286\pm0.004, \label{M-a-RP}
\eeq
in agreement with the estimates given by the Monte Carlo technique results in \cite{Mot-etal:2014a:MNRAS:}, only two other twin HF QPOs models can fit the optical mass limit. First is the so called total precession model, with the frequency identification given by $\nu_{\U}=\nu_{\phi}$ and $\nu_{\L}=\nu_{\theta}-\nu_r$, implying the limits 
\beq
     M = (5.5\pm0.1)~M_{\odot},\quad a = 0.276\pm0.003. \label{M-a-RP1}
\eeq
Second is the resonance epicyclic model demonstrating a beat frequency, with the frequency identification $\nu_{\U}=\nu_{\theta}$ and $\nu_{\L}=\nu_{\theta}-\nu_{\rm r}$, implying the limits  
\beq
     M = (5.1\pm0.1)~M_{\odot},\quad a = 0.274\pm0.003. \label{M-a-RE1}
\eeq
Clearly, all the models predict the black hole spin $a<0.3$. We conclude that, unfortunately, none of the considered geodesic oscillatory models of twin HF QPOs combined with the nodal precession model of LF QPOs can predict values of the black hole spin that could be matched to the predictions of the X-ray spectral measurements, if we assume that the simultaneously measured QPOs occur at the same radius being thus physically related. (For completeness, we have tested also the fitting to the mass estimate $M = (7.02\pm0.22)~M_{\odot}$ -- then the models denoted in \cite{Stu-Kol:2016:ASTRA:} as epicyclic resonance model ER3 and the tidal distortion model TD meet this mass limit for the spin ranges $a=0.44\pm0.02$ and $a=0.46\pm0.04$ that are again in contradiction with the results of both the spin spectral estimates.) 

The controversy of the GRO J1655-40 black hole spin estimates could be solved by using models of twin HF QPOs that are based on frequencies reflecting non-geodesic effects, including thus not only the Kerr spacetime parameters, but also some additional parameters related to the non-geodesic phenomena. We test in the present paper the string loop oscillation model \citep{Stu-Kol:2014:PHYSR4:,Stu-Kol:2015:GRG:} potentially reflecting the influence of tension related to internal toroidal magnetic fields in accretion discs \citep{Cre-Stu:2013:PHYSRE:}. Of course, it is worth to consider also possibilities keeping the assumption of the purely geodesic (gravitational) origin of the oscillatory modes giving rise to the twin HF QPOs and the LF QPO. We discuss this possibility in the following section. 

\section{Geodesic models of QPOs matching the spin limits by spectral measurements} 

In order to keep assumption of validity of purely geodesic models of twin HF QPOs for explanation of observed QPOs in the GRO J1655-40 microquasar and enable possibility to match the black hole spin values predicted by the spectral measurements methods and the mass limit given by the optical measurements, we have to abandon the assumption that the dimensionless radius $x_{\rm HF}$ where the twin HF QPOs occur coincides with the dimensionless radius $x_{\rm LF}$ where the simultaneously observed LF QPO occurs. The HF QPOs and the LF QPO are thus assumed to be physically independent. We still keep assumption that the LF QPO is determined by the relativistic nodal oscillations, but these LF oscillations are not correlated with the twin HF QPOs. Then we can treat the three simultaneously observed QPOs by the frequency relation method proposed in \cite{Stu-Kot-Tor:2013:ASTRA:}, but the version introduced in \cite{Stu-Kol:2016:ASTRA:} has to be slightly modified due to the fact that we do not assume $x_{\rm HF}=x_{\rm LF}$. Of course, we can determine the ratio of these two radii in dependence on the twin HF QPOs model, in the range of allowed values of the black hole spin. We first introduce the geodesic models of QPO and then we use them to fit the QPO data observed in GRO J1655-40 source. 

\subsection{Frequencies of the geodesic epicyclic motion}\label{geneqcond}

In the Kerr spacetimes, circular geodesics exist only in the equatorial plane \citep{Bar-Pre-Teu:1972:ApJ:,Stu:1980:BAC:}. The radial epicyclic frequency $\nu_{\rm r}$, and the vertical epicyclic frequency $\nu_{\theta}$ of the near-circular epicyclic motion are given by the relations  \citep{Kat-Fuk-Min:1998:BHAccDis:,Ste-Vie:1998:ApJ:,Stu-Sche:2012:CLAQG:}
\beq
\label{frequencies}
\nu_{\rm r}^2 = \alpha_\mathrm{r}\,\nu_\mathrm{\phi}^2, 
\quad
\nu_{\theta}^2 = \alpha_\theta\,\nu_\mathrm{\phi}^2,
\eeq
where the orbital (azimuthal) frequency $\nu_\mathrm{\phi}$, sometimes called Keplerian frequency, and the related dimensionless epicyclic frequencies are given by the formulae 
\bea
\nu_{\rm \phi}&=&\frac{1}{2\pi}\left(\frac{c^3}{GM}\right) \frac{1}{ x^{3/2}+a },\nonumber\\
\alpha_{\rm r}&=&1-\frac{6}{x}+ \frac{8a}{x^{3/2}} - \frac{3a^2}{x^{2}} ,
\nonumber\\
\alpha_\theta&=& 1-\frac{4a}{x^{3/2}}+\frac{3a^2}{x^{2}} .
\eea
The dimensionless radius $x = r/r_{\rm g}$ is introduced, where the gravitational radius of the black hole, $r_{\rm g} = GM/c^2$. The radial profiles of the epicyclic frequencies are illustrated in Fig.1. 

Since all the frequencies $\nu_{\phi}, \nu_{\theta}, \nu_{\rm r}$ have the same mass scaling, it is clear that in the geodesic models of twin HF QPOs containing only linear combinations of these frequencies, the frequency ratio of the lower and upper frequencies will be independent of the mass parameter $M$, being dependent only on the spin parameter $a$ \citep{Stu-Kot-Tor:2013:ASTRA:}. This fact enables an effective application of the frequency relation method in the case of the geodesic models of twin HF QPOs. 

\begin{figure*}
\includegraphics[height=\ricpic]{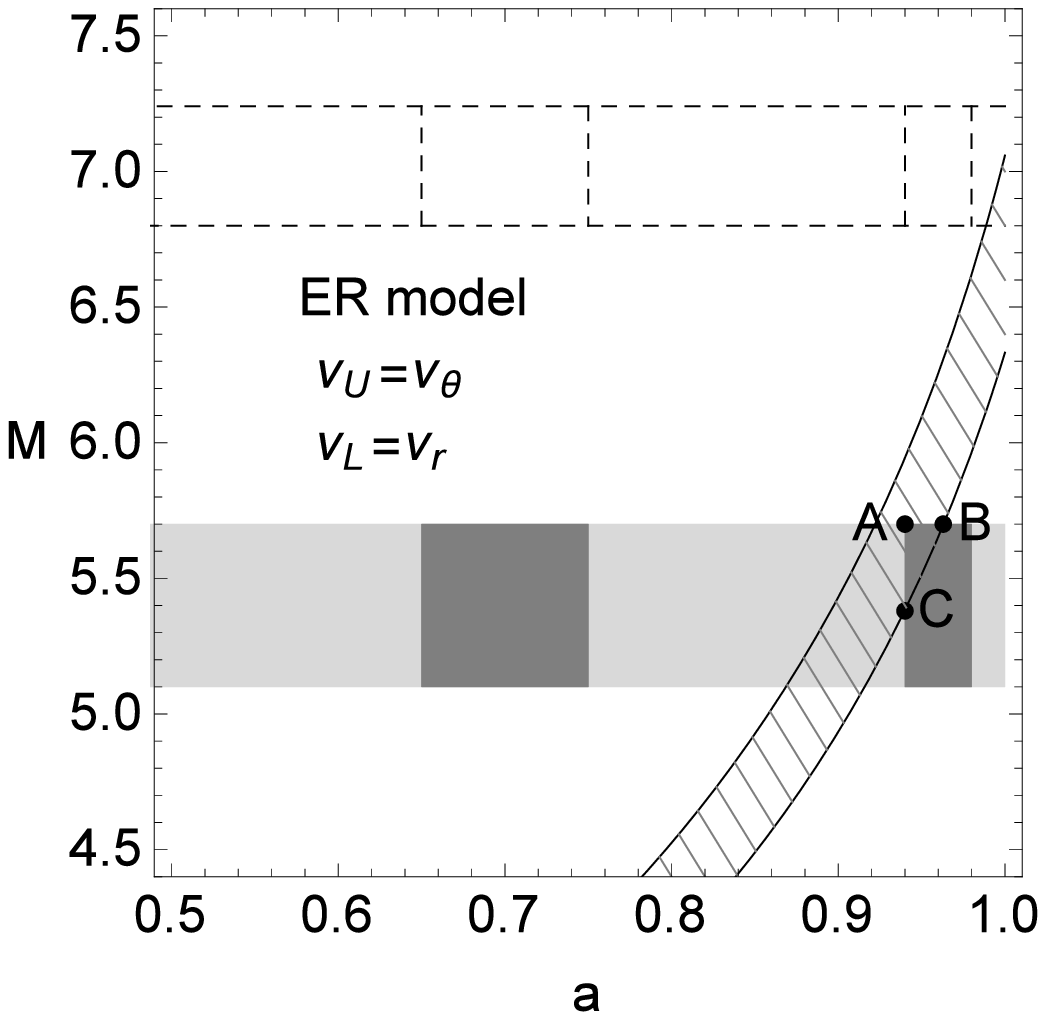}
\includegraphics[height=\ricpic]{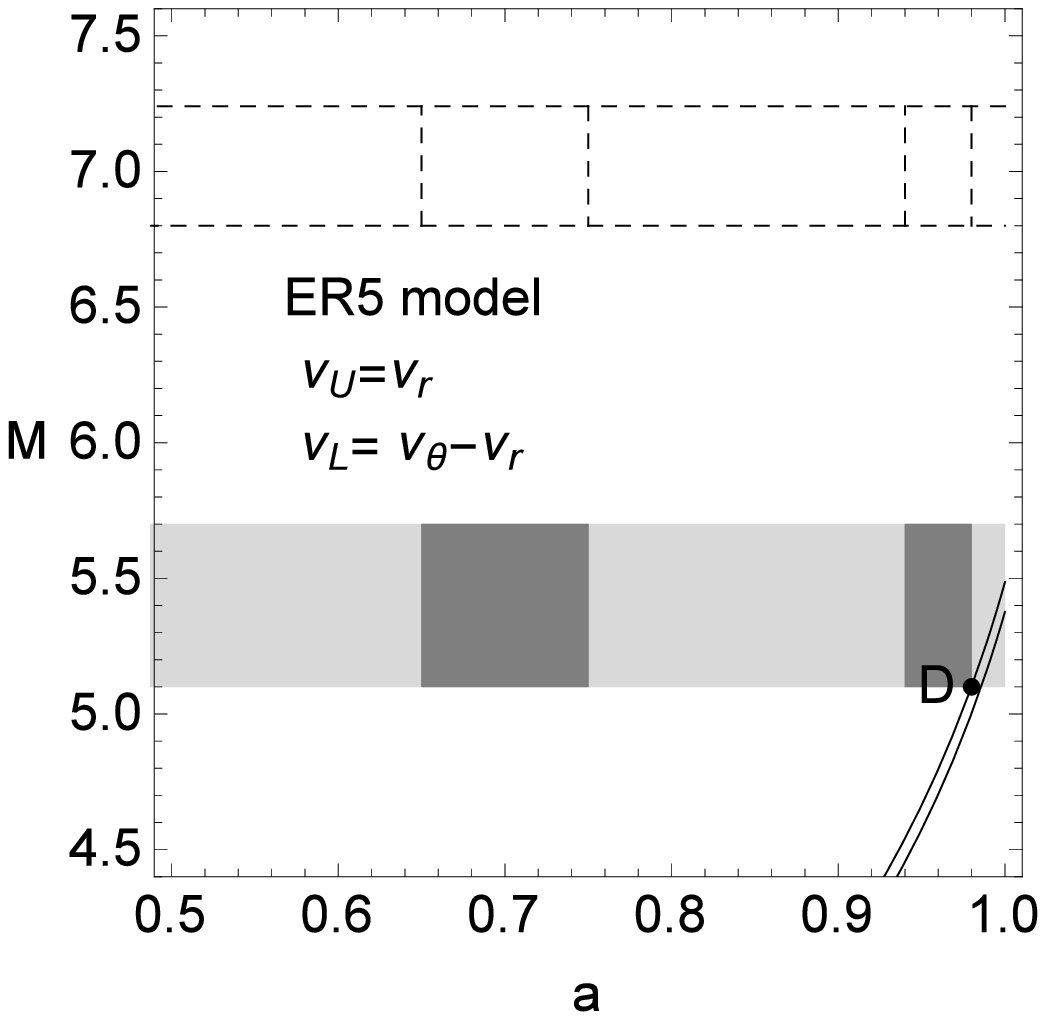}\\
\includegraphics[height=\ricpic]{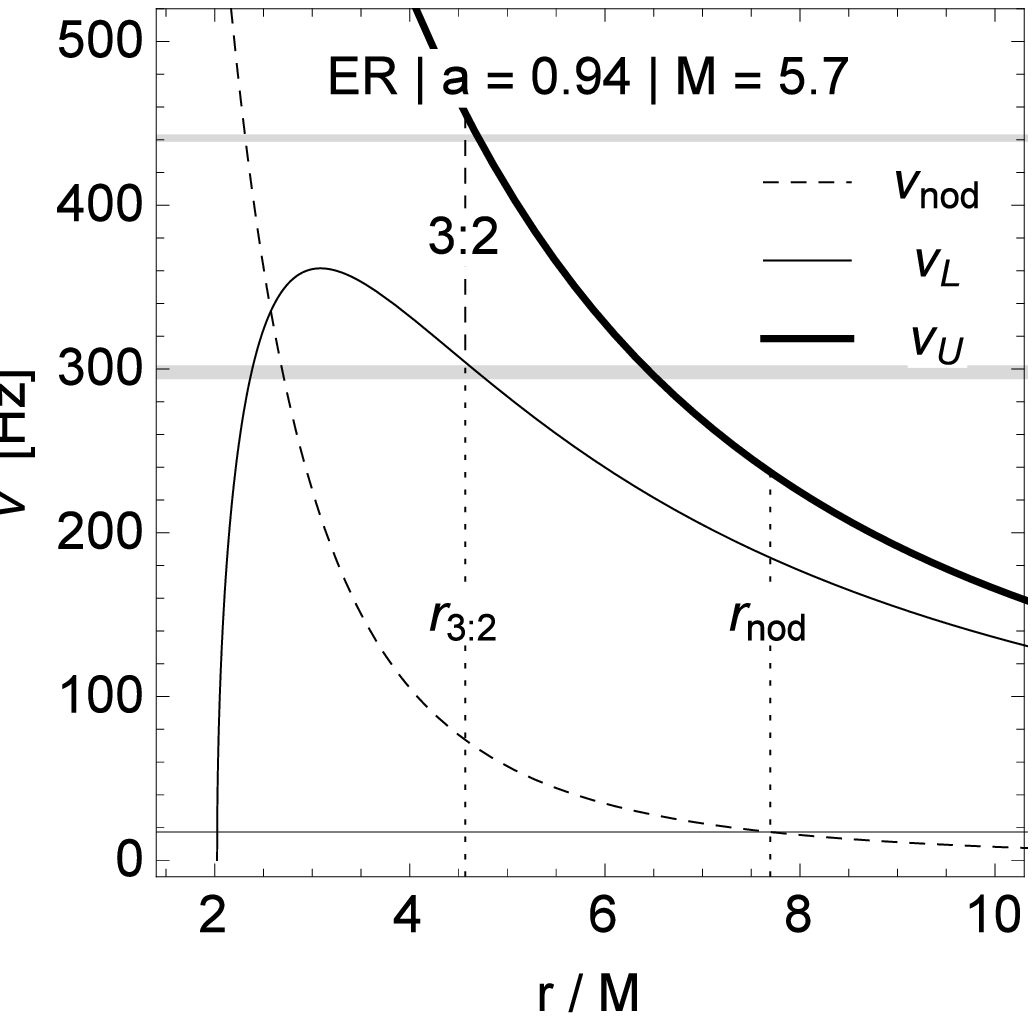}
\includegraphics[height=\ricpic]{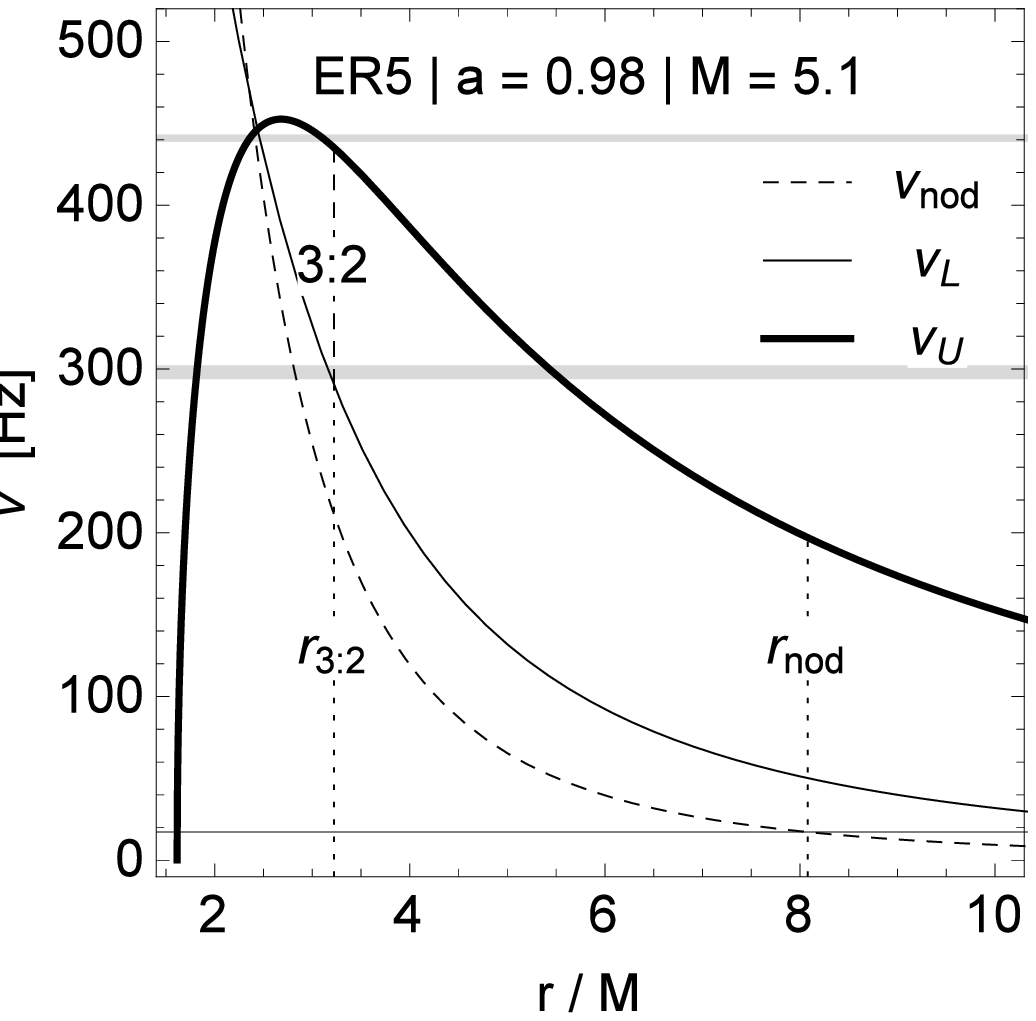}
\caption{
Restrictions on the GRO J1655-40 black hole parameters $M$ and $a$ given by the ER and ER5 models due to the matching of the twin HF QPOs. In the upper row, mass-spin relation due to the geodesic models demonstrates  satisfactory matching to the optical mass limit (shaded area) and the spectral continuum dimensionless spin limits (dark shaded regions). The edges of the cross region are denoted as points A,B,C for the ER model, and the point D for the ER5 model. In the lower row the radial profiles of the upper and lower frequencies in the ER and ER5 models are confronted with the nodal precession frequency radial profile, and the radii where the twin HF QPOs and the LF QPO occur are given. 
\label{fig1}
}
\end{figure*}

\subsection{Models based on the epicyclic geodesic motion}

The geodesic models of twin HF QPOs can be separated into three classes: the hot spot models (the RP model and its variants \citep{Ste-Vie:1999:PHYSRL:,Stu-Kot-Tor:2013:ASTRA:}, the tidal precession model \citep{Kos-etal:2009:ASTRA:}), resonance models \citep{Tor-etal:2005:ASTRA:,Stu-Kot-Tor:2011:ASTRA:} and disc oscillation (discoseismic) models \citep{Rez-etal:2003:MNRAS:,Mon-Zan:2012:MNRAS:}. These models were applied to match the twin HF QPOs and the LF QPO in the microquasar GRO J1655-40 \citep{Stu-Kol:2016:ASTRA:}. We have tested in the present study all the models considered in \cite{Stu-Kol:2016:ASTRA:}. Here we briefly summarize properties of the resonance models that were shown to be the only sucessful models in matching at least one of the spin limits predicted by the spectral measurements, along with the mass limits implied by the dynamical restrictions due to optical measurements. 

The LF QPO remains related to the relativistic nodal (Lense-Thirring) precession with frequency $\nu_{nod}=\nu_{\phi}-\nu_{\theta}$. The observations demonstrate that LF QPOs given by the Lense-Thirring effect can occur independently on the twin HF QPOs, and only exceptionally their simultaneous occurrence is observed \citep{Mot-etal:2014a:MNRAS:}. Therefore, we are approved to use the assumption of the twin HF QPOs and the LF QPO simultaneously observed, but physically uncorrelated. 

The epicyclic resonance (ER) models \citep{Abr-Klu:2001:AA:,Tor-etal:2005:ASTRA:} consider resonance of axisymmetric oscillation modes of accretion discs. The accretion discs can be geometrically thin, having the geodetical (Keplerian) profile of angular velocity \citep{Nov-Tho:1973:BlaHol:,Pag-Tho:1974:ApJ:}, or toroidal, geometrically thick, with angular velocity profile determined by pressure gradients \citep{Koz-etal:1977:ASTRA:,Abr-etal:1978:ASTRA:,Stu-etal:2009:CLAQG:}. Frequencies of the disc oscillations are related to the orbital and epicyclic frequencies of the circular geodesics for both the Keplerian discs \citep{Kat-Fuk-Min:1998:BHAccDis:,Kat:2004:PASJ:,Now-Leh:1998:TBHAD:} and slender tori \citep{Rez-etal:2003:MNRAS:,Mon-Zan:2012:MNRAS:}. 
The resonance can be of two kinds. 
The internal, parametric resonance of the radial and vertical epicyclic oscillatory modes, representing the basical resonance epicyclic model, governed by the Mathieu equation, which predicts the strongest resonant phenomena for the frequency ratio $3:2$ \citep{Lan-Lif:1969:Mech:,Nay-Moo:1979:NonOscilations:,Stu-Kot-Tor:2013:ASTRA:}. 
The forced non-linear resonance admits presence of combinational (beat) frequencies in the resonant solutions \citep{Nay-Moo:1979:NonOscilations:}. \footnote{Variants of the resonance model with beat frequencies are presented in \cite{Stu-Kol:2016:ASTRA:} -- all the variants were considered in the present study.} Of course, the resonant phenomena could be relevant also in the framework of the hot spot models \citep{Stu-Kot-Tor:2012:ACTA:,Stu-etal:2015:ACTA:}. 

While in the ER models the oscillatory modes of the accretion disc are assumed axisymmetric, in the warped disc (WD) oscillation model using the inertial-acoustic modes and the so called g-modes of thin discs oscillations the oscillations are assumed non-axisymmetric \citep{Kat:2004:PASJ:,Kat:2008:PASJ:}. 

The parametric resonance admits slight scatter of the resonant frequencies, i.e., this kind of resonance can occur while the oscillating modes in resonance have frequency ratio slightly different from the exact rational ratio; width of the resonance scatter decreases with increasing order of the resonance \citep{Lan-Lif:1969:Mech:}. For forced resonances, scatter of frequency ratio from the rational ratio is governed by non-linear effects \citep{Nay-Moo:1979:NonOscilations:}. Therefore, we consider as relevant all the frequency ratios given by the measured HF QPO frequencies with their errors. 

\begin{figure*}
\includegraphics[height=\ricpic]{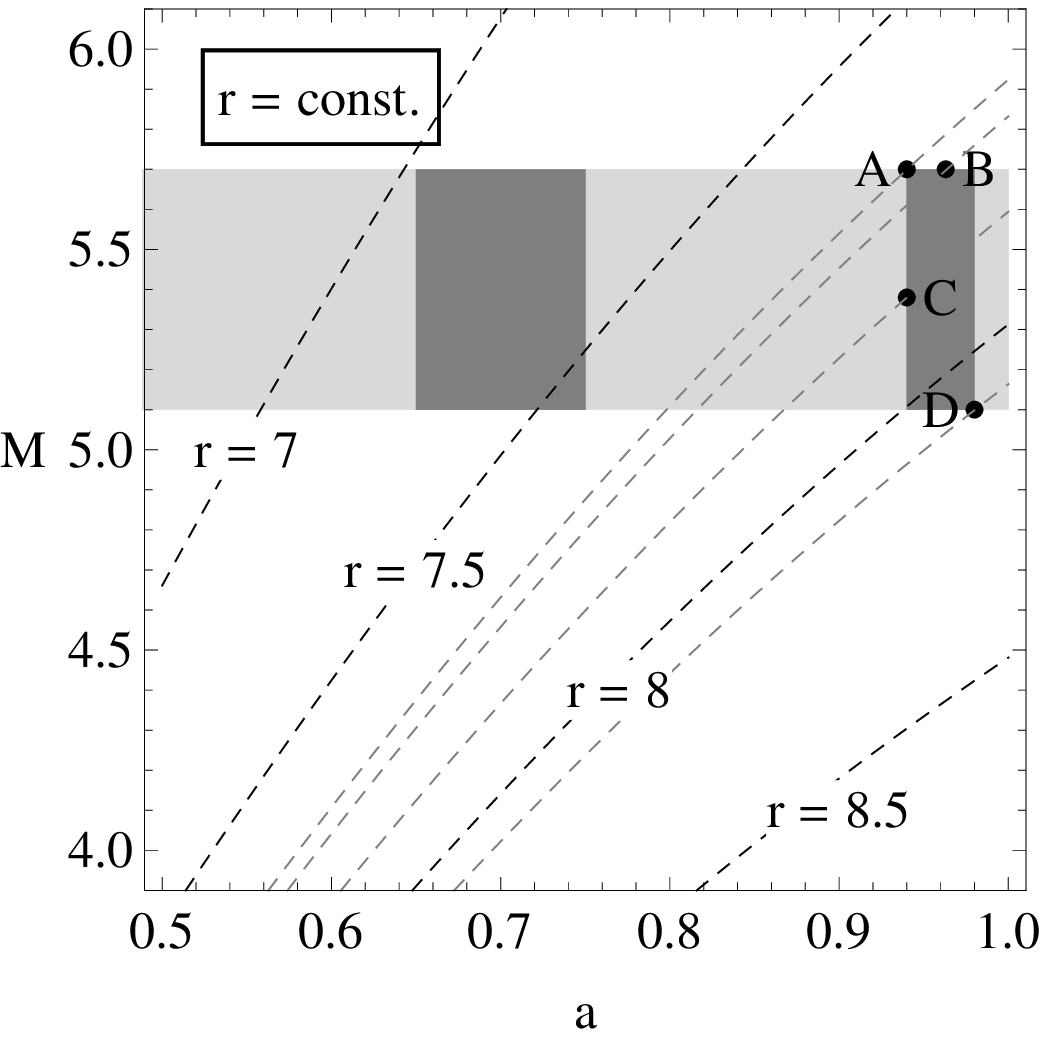}
\includegraphics[height=\ricpic]{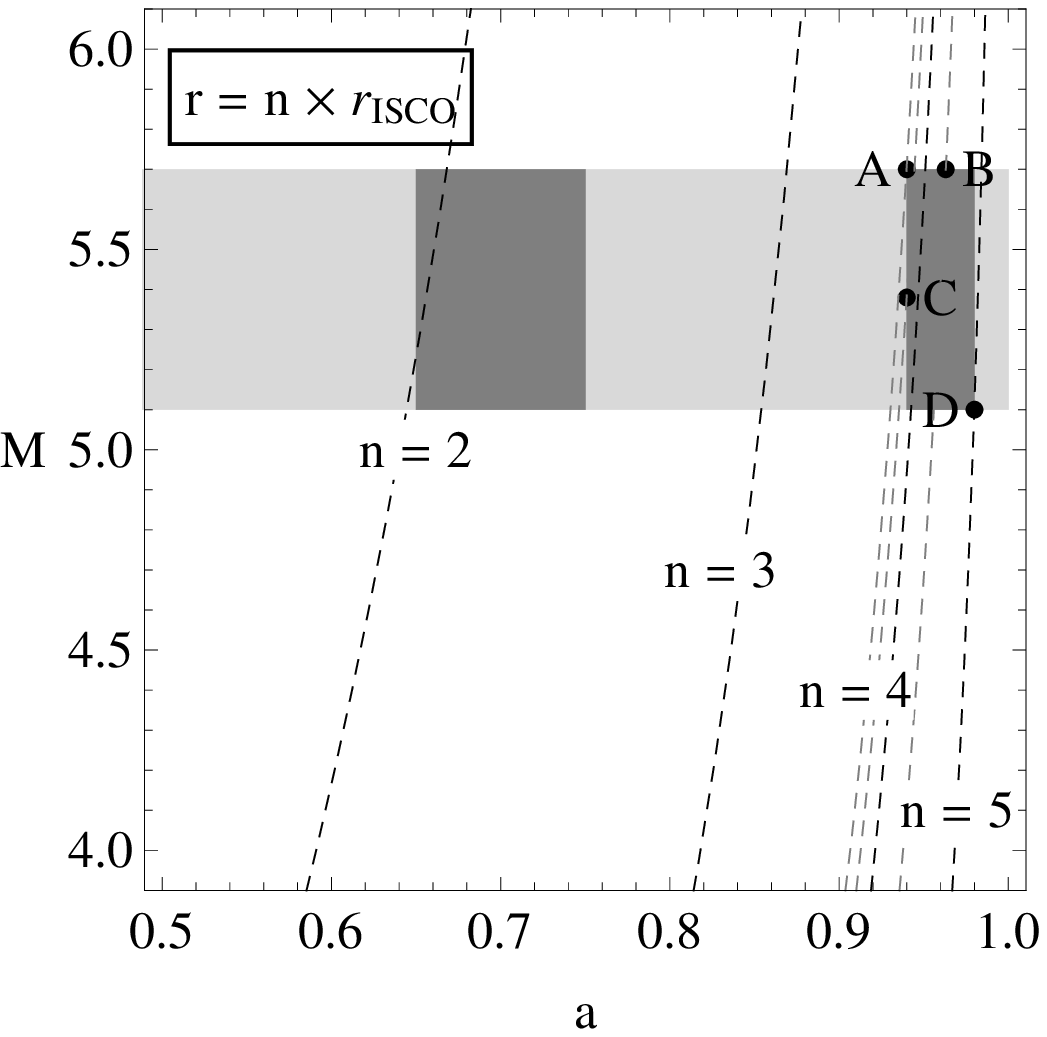}\\
\includegraphics[height=\ricpic]{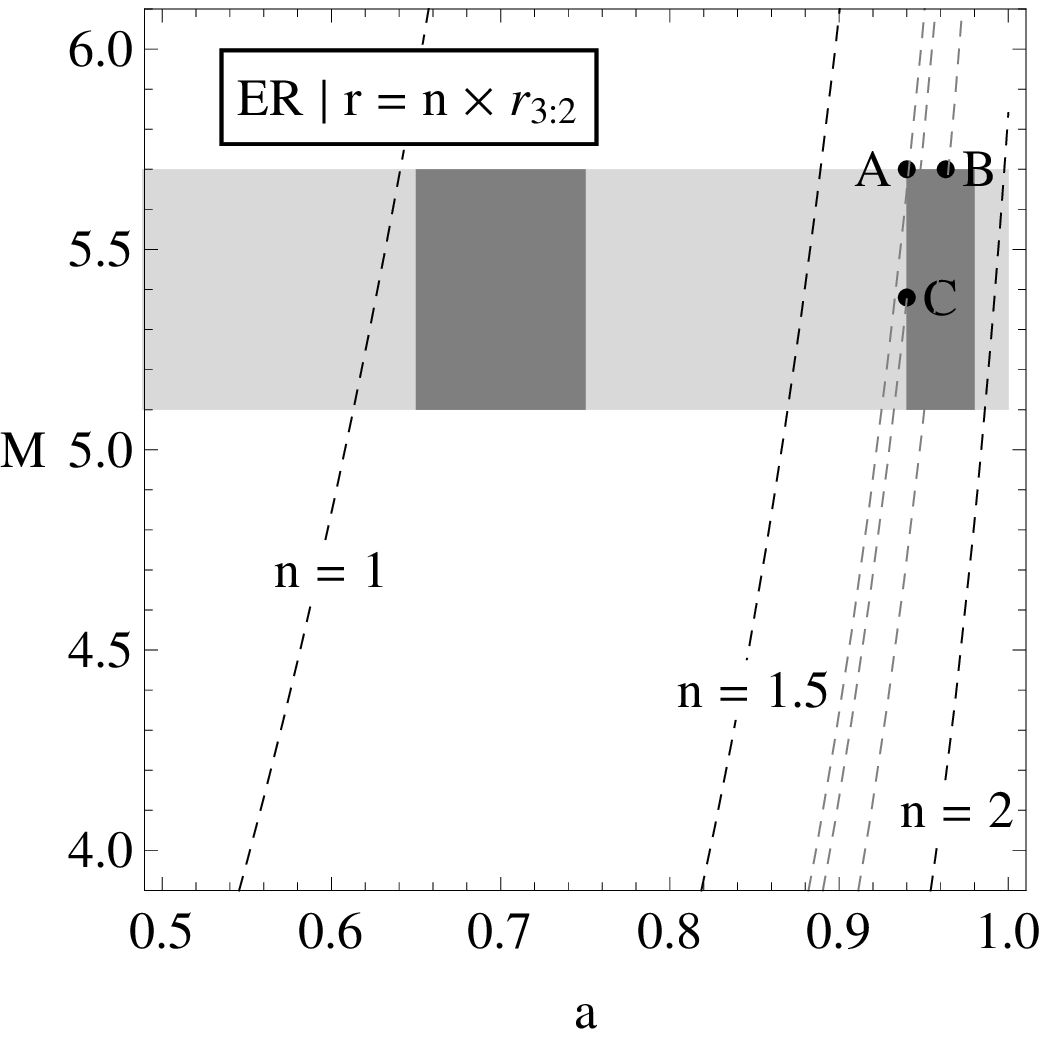}
\includegraphics[height=\ricpic]{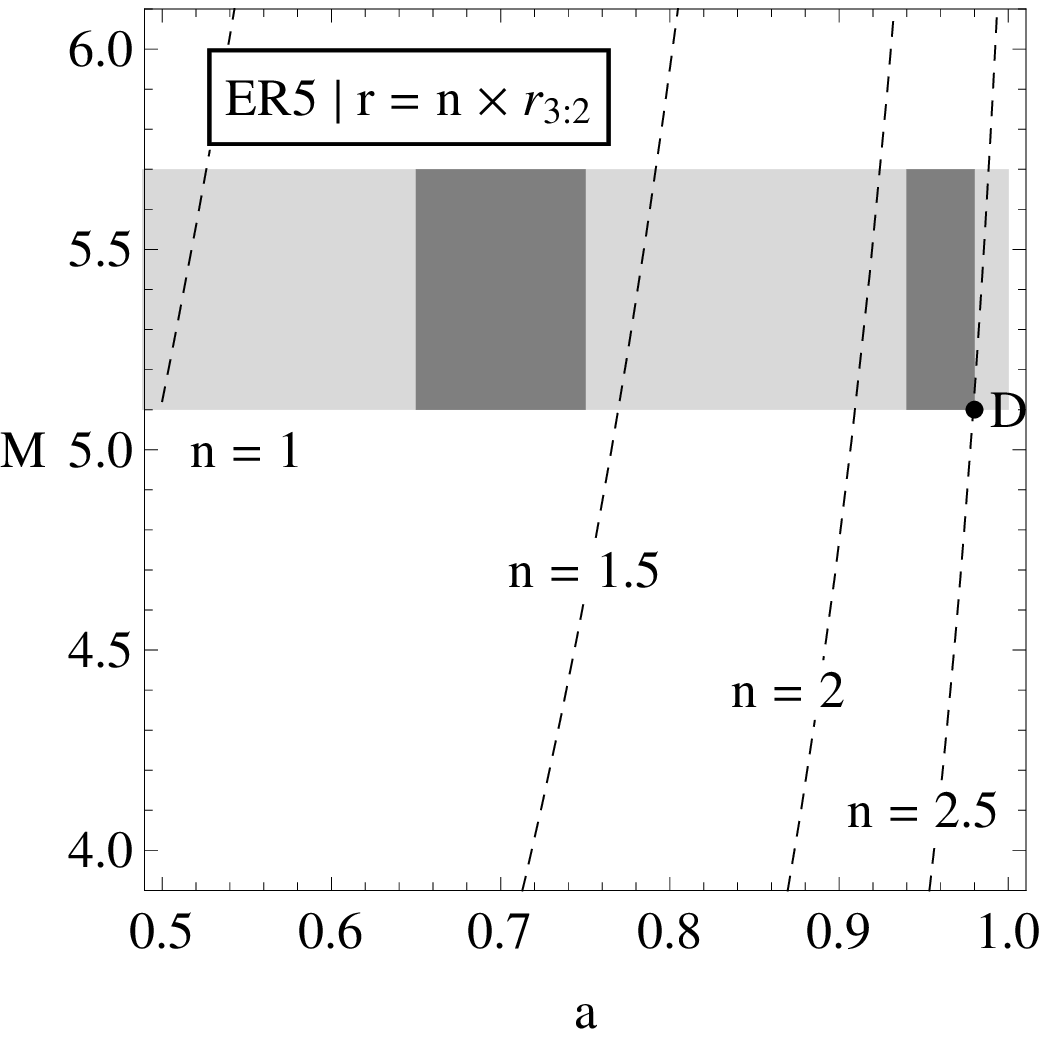}
\caption{
Restrictions on the parameters $M$ and $a$ given the relativistic nodal precession model related to the LF QPO observed simultaneously with the twin HF QPOs in the microquasar GRO J1655-40. 
\label{fig2}
}
\end{figure*}

\begin{table*}
\begin{center}
\begin{tabular}{l |  c  c  | c c | c c}
\hline
 & a & M & $r_{3:2}$ & $r_{\rm nod}$ & $r_{\rm nod}/r_{\rm ms}$ & $r_{\rm nod}/r_{3:2}$ \\
\hline \hline
ER (point A) & 0.94 & 5.7 & 4.73 &  7.68--7.71 & 3.79--3.81 & 1.62--1.63 \\
ER (point B) & 0.96 & 5.7 & 4.71 &  7.72--7.75 & 4.26--4.28 & 1.64--1.65 \\
ER (point C) & 0.94 & 5.38 & 4.96 & 7.84--7.87 & 3.87--3.89 & 1.58--1.59 \\
\hline
ER5 (point D) & 0.98 & 5.1 & 3.14 &  8.06--8.10 & 5.00--5.02 & 2.56--2.58 \\
\hline
\end{tabular}
\caption{
Restrictions on the parameters $M$ and $a$ of the black hole in the microquasar GRO~J1655-40 given by the geodesic oscillation models of twin HF QPOs. 
\label{tab1}
} 
\end{center}
\end{table*}

\subsection{Matching the observed QPO frequencies to the geodesic oscillation models}

Considering a twin HF QPOs geodesic model we determine for a given frequency ratio parameter $p$ 
the frequency relations 
\beq
     a = a^{\nu_\U (\phi,r,\theta)/\nu_\L (\phi,r,\theta)}(x,p)
\eeq and (by a numerical procedure) the corresponding frequency relations governing the radius of occurrence
\beq 
     x = x^{\nu_\U(\phi,r,\theta)/\nu_\L(\phi,r,\theta)}(a,p) ;  
\eeq 
the twin oscillations with the upper (lower) frequency $\nu_\U(\phi,r,\theta)$ ($\nu_\L(\phi,r,\theta)$) are determined by the concrete geodesic model. The radius of occurrence have to satisfy the condition $x^{\nu_\U/\nu_L}(a,p) \geq x_{\rm ms}(a)$, where radius of the marginally stable orbit $x_{\mathrm{ms}}(a)$ is implicitly given by \citep{Bar-Pre-Teu:1972:ApJ:,Stu-Kot-Tor:2013:ASTRA:}
\beq
a=a_{\rm ms}\equiv\frac{\sqrt{x}}{3}\left(4-\sqrt{3x-2}\right).
\eeq

Since the assumption of the coincidence of the radii where the twin HF QPOs and the simultaneously observed LF QPO were observed is abandoned now, we have to use the frequency relation technique developed in \cite{Stu-Kol:2016:ASTRA:} in slightly modified form where we consider separately the fitting of the twin HF QPOs giving the dependence $M_{\rm HF}(a,p)$ and the dependence $M_{\rm LF}(a,x)$ for an arbitrarily fixed radius $x$ where the LF QPO can occur. 

We have tested all the geodesic models of the twin HF QPOs studied in \cite{Stu-Kol:2016:ASTRA:}, demonstrating that none of the models can be in accord with the limits implied by the spectral continuum measurements ($0.65 < a < 0.75$). On the other hand, two geodesic models of the twin HF QPOs are in (partial) agreement with the Fe-line spectral measurements ($0.94 < a < 0.98$). We focus attention on these two satisfactory models. We give detailed description of the method for one of the models, namely the ER model. The method of matching the observational data constitutes from the following successive steps. 

\subsubsection{Matching procedure for the ER model}

First, we determine the frequency ratio interval of the lower and upper centroid frequencies of the measured twin HF QPOs with the related errors. Assume that the interval reads $p_1 < p < p_2$. For each ratio from the interval, we have to find the $M_{\rm HF}(a,p)$ relation. To find the spin-radius and mass-spin relations with errors corresponding to the measurement errors of the twin HF QPOs frequencies, it is enough to make the calculations for the frequency ratio parameters $p_1$ and $p_2$ -- these errors represent the maximal errors, as opposed to the statistical errors \citep{Stu-Kol:2016:ASTRA:}. 

Second, we use the frequency relation of the ER model 
\bea
 a&=&a^{\mathrm{\theta/r}}(x,p) \nonumber\\
 &\equiv& \frac{\sqrt{x}}{3(p+1)}\Big[2(p+2) \nonumber \\
 && -\sqrt{(1-p)[3x(p+1)-2(2p+1)]}\Big], 
\eea
and determine by numerical procedure the inverse frequency relation for the radius where the twin HF QPOs have to occur $x^{\theta/r,}(a;p)$; the spin parameter is assumed in the interval $0 \leq a \leq 1$. 

Third, using the relation 
\beq
\nu_{\rm U} = \nu_{\theta} = \frac{1}{2\pi}\left(\frac{c^3}{GM}\right)
\frac{(1-4\,a\,x^{-3/2}+3a^2\,x^{-2})^{1/2}}{(x^{3/2}+a)},
\eeq
where for a given spin $a$ we apply the numerically determined relation $x=x^{\theta/r,}(a;p)$, we obtain the mass-spin relation $M_{\rm HF}^{\theta/r}(a,p)$. We construct the HF mass-spin relation for the limiting values of $p=p_1, p=p_2$ reflecting the error in determining the mass-spin relation connected to the measurement errors of twin HF QPOs. The mass-spin relation $M_{\rm HF}^{\theta/r}(a,p)$ related to the ER model of twin HF QPOs is illustrated in Fig.1 (left column). We can see that the function $M_{\rm HF}^{\theta/r}(a,p)$ really satisfies simultaneously the optical mass limit and the spin limit of the Fe-line spectral measurements, but not for the whole intervals given by the optical measurements of mass and the Fe-line limits on spin. In fact the ER model introduces additional restrictions: $M>5.4M_{\odot}$ and $a<0.96$. 

Fourth, the frequency of the relativistic nodal precession in the Kerr geometry reads \citep{Ste-Vie:1998:ApJ:} 
\beq
   \nu_{\rm nod}(x;M,a) = \frac{1}{2\pi} \frac{c^3}{GM} \frac{[1 - (1 - \frac{4a}{x^{3/2}} + \frac{3a^2}{x^2})^{1/2}]}{ x^{3/2}+a } .
\eeq 
Under the assumption of the coincidence of the radii of occurrence of the twin HF QPOs and the LF QPO, we used the condition $x=x_{\rm HF}(a,p)$ \citep{Stu-Kol:2016:ASTRA:}. Here we fix the radius $x={\rm const}.$ and use the condition 
\beq
        \nu_{\rm LF} = \nu_{\rm nod}(x,M,a)
\eeq
in order to determine the LF mass-spin relation $M_{\rm LF}(a,x)$ governed by the relativistic nodal precession. We numerically determine the radii $x$ for which the mass-spin relations $M_{\rm LF}(a,x)$ cross the region of the mass-spin parameter space corresponding to the optical measurement limits on the mass and the spectral limits on the spin of the GRO J1655-40 black hole. This procedure is now independent of the geodesic models of the twin HF QPOs. The results given in units of gravitational mass $M$, or the radius $x_{\rm ms}$ are represented in Fig.2. Of course, they can be expressed also in terms of the corresponding $x_{\rm HF}(a,p)$. 

\subsubsection{Matching procedure for the ER5 model with beat frequency}

We have found one model with beat frequencies that meets the optical limits on the black hole mass and the Fe-line limits on the black hole spin. The identification of the ER5 model \footnote{We keep the notation of the models as presented in \cite{Stu-Kol:2016:ASTRA:}} with beat frequencies reads 
\beq
      \nu_{\rm L} = \nu_{\theta} - \nu_r , \quad \nu_{\rm U} = \nu_r . 
\eeq
The frequency relation of the ER5 model reads
\begin{equation}
a = a^{\mathrm{r/(\theta-r)}}(x,p) = a^{\mathrm{\theta/r}}(x,p^{\rm \prime})
\end{equation}
where 
\begin{equation}
  p^{\rm \prime} = \frac{1}{(1 + \sqrt{p})^2} . 
\end{equation}

Following the first three steps of the procedure of determination of the mass-spin relation presented above, we arrive to $M_{\rm HF}^{r/(\theta-r)}(a,p)$. The resulting curve is presented in Fig.1. We can see that the predictions of the RE5 model only touch the mass-spin region given by the optical measurement and the Fe-line fittings at the point $M=5.1M_{\odot}$ and $a=0.98$. 

The resulting regions of the mass-spin parameter space allowed by the ER and ER1 geodesic models are presented in Table 1, along with the radii $x_{\rm HF}$ and $x_{\rm LF}$.
{Note that none of the considered twin HF QPO models is in agreement with the mass estimate $M = (7.02\pm0.22)~M_{\odot}$.} 

{

\section{String loop oscillation model}

The current-carrying string loops \citep{Lar:1993:CLAQG:,Lar:1994:CLAQG:,Jac-Sot:2009:PHYSR4:,Kol-Stu:2010:PHYSR4:} represent one of the models based on non-geodesic phenomena that could reflect plasma exhibiting a string-like behavior due to dynamics of the magnetic field lines \citep{Sem-Dya-Pun:2004:Sci:,Chri-Hin:1999:PhRvD:}, or due to the thin flux tubes of magnetized plasma described as 1D strings \citep{Sem-Ber:1990:ASS:,Cre-Stu:2013:PHYSRE:,Cre-Stu-Tes:2013:PhysPlasm:,Kov:2013:EPJP:,Kol-Stu-Tur:2015:CLAQG:,Tur-Stu-Kol:2016:PHYSR4:}. The high-energy string loops can serve as a model of formation and collimation of ultra-relativistic jets in the field of black holes or naked singularities located in active galactic nuclei or Galactic microquasars \citep{Stu-Kol:2012:PHYSR4:,Stu:1983:BULAI:,Stu-Hle:1999:PHYSR4:,Stu-Kol:2012:JCAP:,Kol-Stu:2013:PHYSR4:,Tur-etal:2014:PHYSR4:}, while the low-energy string loops can serve as a model of twin HF QPOs occuring in accretion discs orbiting black holes or neutron stars \citep{Stu-Kol:2012:JCAP:,Stu-Kol:2014:PHYSR4:,Stu-Kol:2015:MNRAS:,Stu-Kol:2015:GRG:}

\begin{table*}
\begin{center}
\begin{tabular}{l l l l l l | l l}
\hline
model & $\nu_\U$ & $\nu_\L$  & $M_{\rm min}$--$M_{\rm max}/M_{\odot}$ & $x_{\rm min}$--$x_{\rm max}$ & $a$ & $\omega$ & $x$ \\
\hline \hline
 string loop 3:2 & $\nu_\theta$ & $\nu_r$ 		 & 5.4--9.9 & 1.7--6.6 & 0.79--1 & $(-1,-0.49)$ & $5.7\pm0.8$ \\
 string loop 2:3 & $\nu_r$ 			& $\nu_\theta$ & 4.9--9.4 & 3.7--8.8 & 0.31--1 & $(-1,-0.12)$ & $8.1\pm0.7$ \\
\hline
\end{tabular}
\caption{
Restrictions on the parameters $M$ and $a$ of the black hole in the microquasar GRO~J1655-40 given by the string loop oscillation model of twin HF QPOs. 
\label{tab2}
} 
\end{center}
\end{table*}

\begin{figure*}
\includegraphics[height=\ricpic]{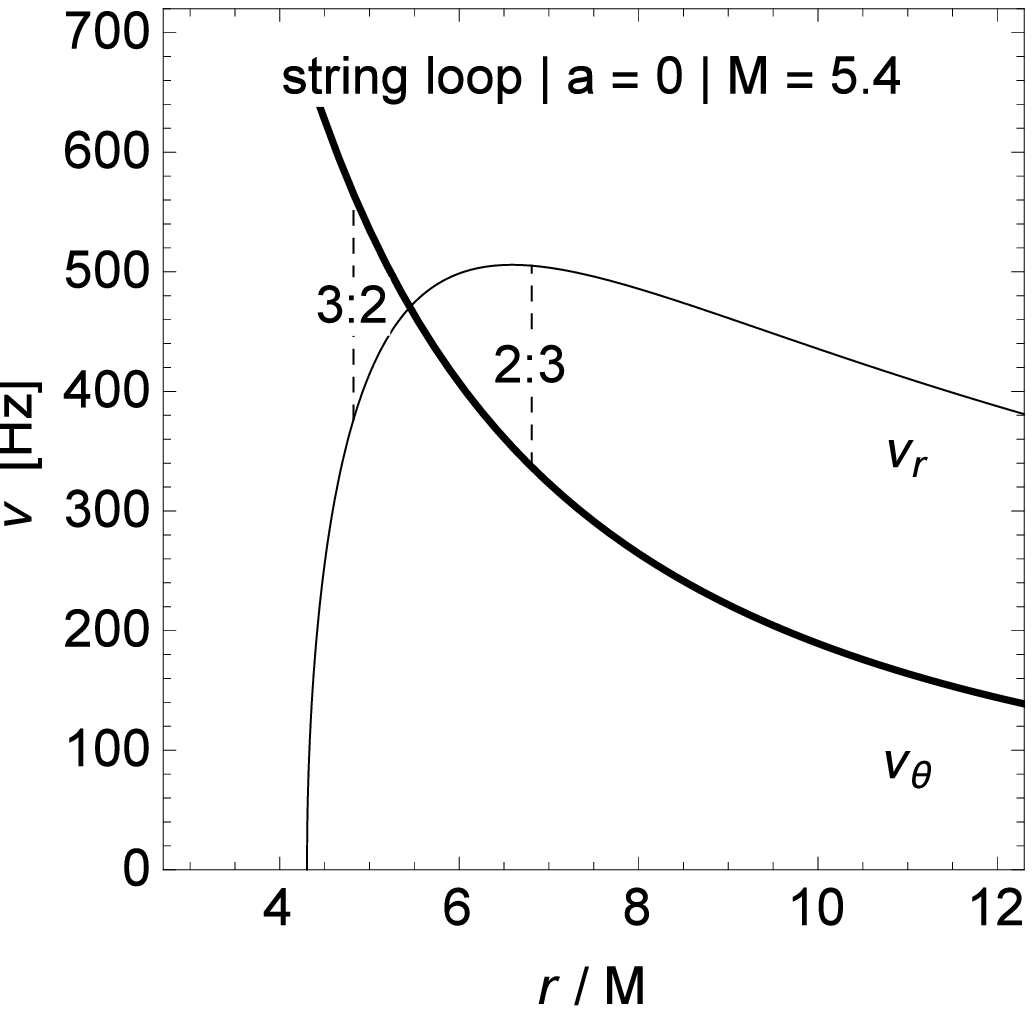}
\includegraphics[height=\ricpic]{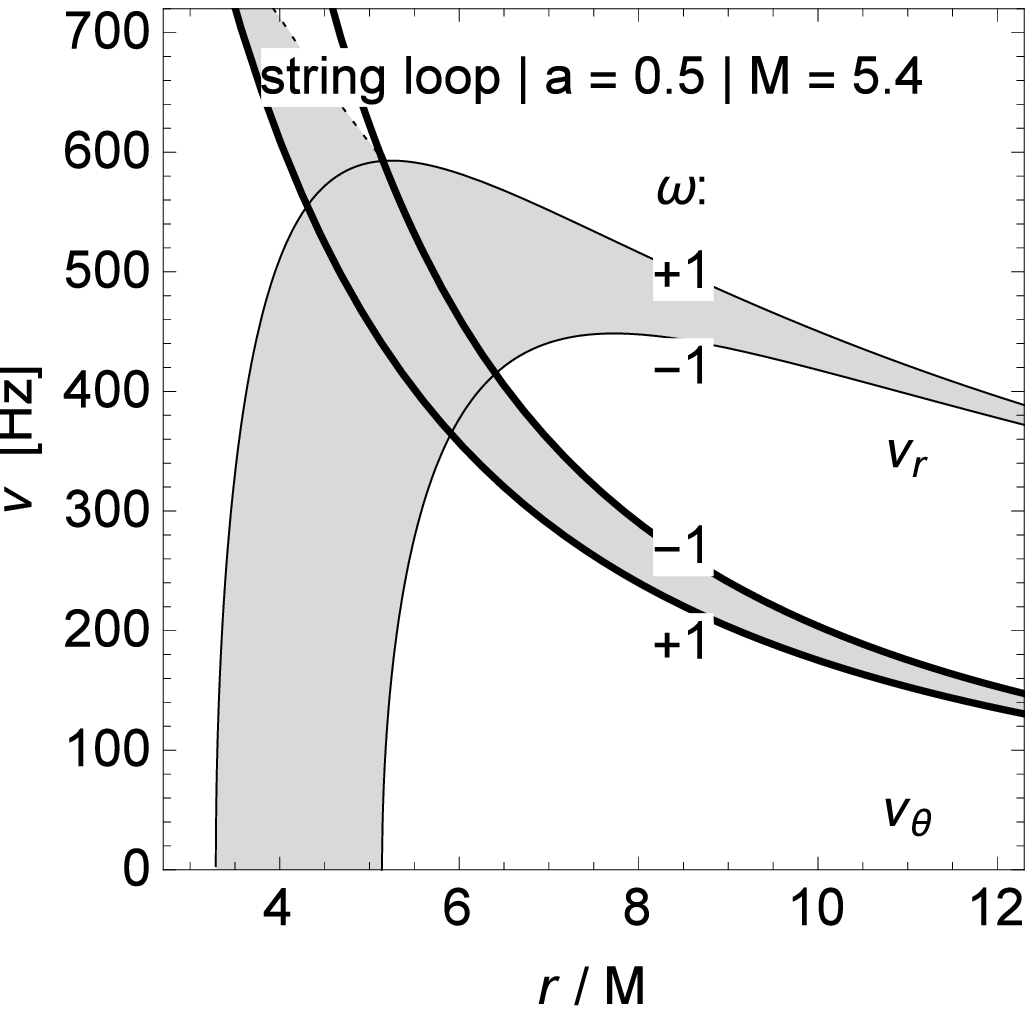}\\
\includegraphics[height=\ricpic]{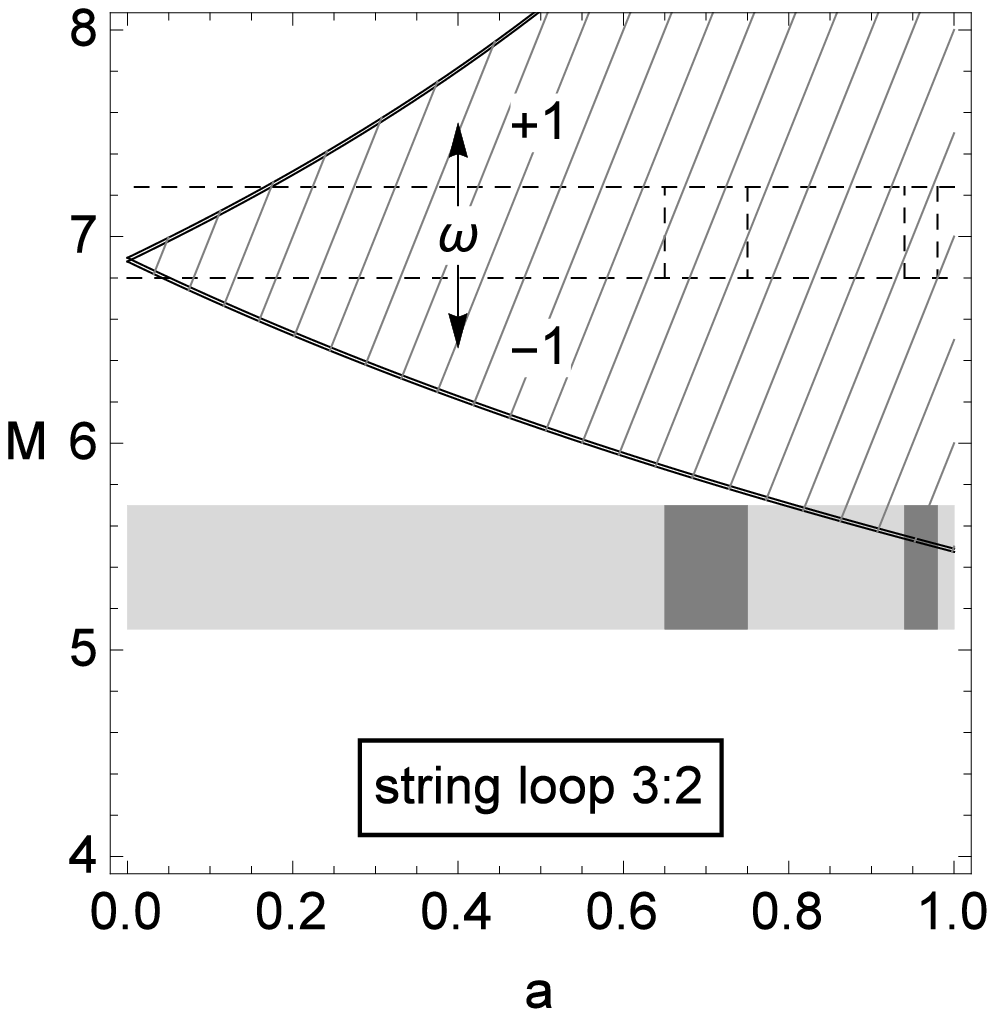}
\includegraphics[height=\ricpic]{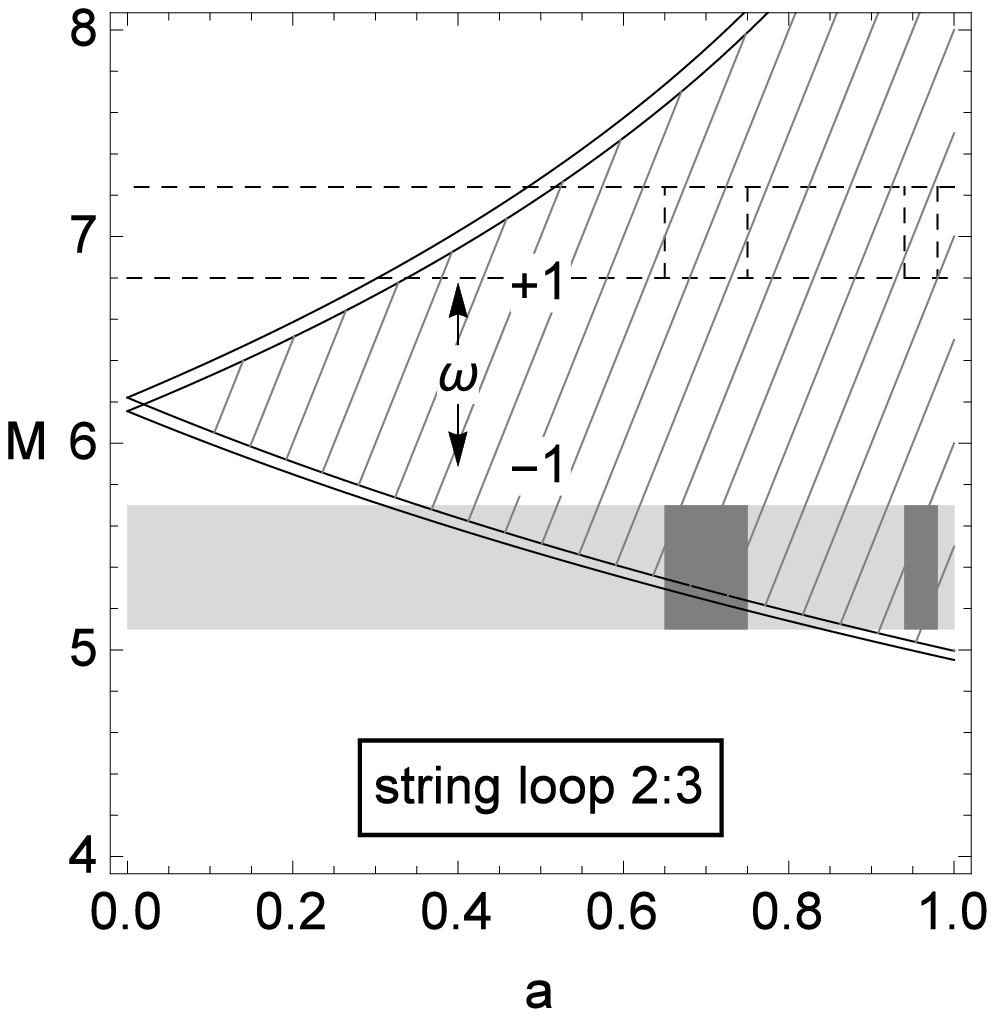}
\caption{
Restrictions on the parameters $M$ and $a$ given by {\it string loop} model for the M82~X-1 source (lower row). 
Solid lines are given by $3\nu_{\rm r}=2\nu_{\rm \theta}$ and $2\nu_{\rm r}=3\nu_{\rm \theta}$ HF~QPOs resonance - hatched  area covers whole $\omega\in\langle-1,1\rangle$ range. Shaded is the mass region limited by the optical measurements. We also give examples of the radial profiles of the frequencies of the radial and vertical harmonic oscillatory modes of the string loops (upper row). 
\label{figLOOP}
}
\end{figure*}

\subsection{Frequency of the string-loop radial and vertical oscillatory modes}

Dynamics of the axisymmetric string loops in the axisymmetric Kerr geometry is governed by the parameters of the gravitational field, the energy parameter $E$ of the string loop, and the two parameters, $J, \omega$, governing the combined effects of the string tension and its angular momentum \citep{Kol-Stu:2013:PHYSR4:,Stu-Kol:2014:PHYSR4:}. Small harmonic or quasi-harmonic oscillations of the string loops can occur around stable equilibrium positions in the equatorial plane of the Kerr geometry. For the radial and latitudinal (vertical) harmonic oscillatory string loop motion in the Kerr spacetimes, the frequencies related to distant observers are given by \citep{Stu-Kol:2014:PHYSR4:} 
\beq
      \nu_{r} = \frac{c^3}{2\pi GM} \, \Omega_{r}, \quad \nu_{\theta} = \frac{c^3}{2\pi GM} \, \Omega_{\theta};
\eeq
the dimensionless angular frequencies read
\begin{widetext} 
\bea
 \Omega^2_{\mir}(r) &=& \frac{ J_{\rm E(ex)} \, \left(2 a \omega \sqrt{\Delta } \left(a^2+3 r^2\right)+\left(\omega ^2+1\right) \left(a^2 r^3 -a^2 \Delta  +r^5-2 r^4\right)\right)}{2 r \left(a^2
   (r+2)+r^3\right)^2 \left(2a\omega  \left(a^2+3r^2\right)+\sqrt{\Delta } \left(\omega ^2+1\right) \left(r^3-a^2\right)\right)^2} , \\
 \Omega^2_{\mit}(r) &=& 
\frac{2 a \omega \sqrt{\Delta }  \left(2 a^2 -3 a^2 r -3 r^3\right)+\left(\omega ^2+1\right) \left(a^4 (3 r-2)+2 a^2 (2 r-3) r^2+r^5\right)}
{ r^2 \left(a^2 (r+2)+r^3\right) \left(2a\omega \left(a^2+3r^2\right) \Delta^{-1/2} + \left(\omega ^2+1\right) \left(r^3-a^2\right)\right)}, 
\eea
where 
\bea
 J_{\rm E(ex)} (r) &\equiv& H \left(\omega ^2+1\right) (r-1) \left( 6 a^2 r -3 a^2 r^2 -6 a^2 -5 r^4+ 12 r^3 \right) \nn \\
 && + \left(\omega ^2+1\right) \left[2 F (a^2+3r^2) (1-r) - F H \right] + 8 a \omega \sqrt{\Delta} (r-1) (a^2+3r^2)^2  \nn \\
 && + 4 a \omega \Delta^{-1/2} H \left[ (a^2+3r^2) \left(\Delta -(r-1)^2\right) -6 r \Delta (r-1) \right]  \label{JEex} 
\eea
and  
\bea
 H(r;a) &=& a^2 (r+2) +r^3, \\
 F(r;a) &=& (r-3) r^4 -2 a^4+a^2 r \left(r^2-3 r+6\right). \label{Ffce}
\eea
\end{widetext}
The zero points of the function $J_{\rm E(ex)}(r;a,\omega)$ determine the marginally stable equilibrium positions of the string loops; at the zero points the frequency of the radial oscillatory mode of the string loops vanishes - for details see \cite{Kol-Stu:2013:PHYSR4:,Stu-Kol:2014:PHYSR4:}. The radial profiles of the frequencies $\nu_{\theta}$ and $\nu_r$ of the string loop harmonic oscillations are demonstrated in Figure \ref{figLOOP}.

\subsection{Fitting the twin HF QPO frequencies}

We assume resonance phenomena in the string loop oscillatory motion, governed by the Kolmogorov-Arnold-Moser theory \citep{Arnold:1978:book:}, as the 3:2 frequency ratio is observed at the twin HF QPOs observed in the GRO~J1655-40 microquasar. We directly identify the observed frequencies $\nu_{\rm U}, \nu_{\rm L}$ with the $\nu_{\mit}, \nu_{\mir}$ or $\nu_{\mir}, \nu_{\mit}$ frequencies. In the case of the string loop oscillation model the additional nodal frequency model can be attributed to a physically independent relativistic nodal precession of a hot spot related to the LF QPO. \footnote{For possibility to obtain a low frequency string loop oscillations see \cite{Stu-Kol:2015:GRG:}}. 

The fitting of the string loop oscillation frequencies to the observed frequencies introduced in \cite{Stu-Kol:2014:PHYSR4:} will be used here. The string loop oscillation model implies a triangular limit on the spacetime parameters $M,a$  -- see Figure \ref{figLOOP}. \footnote{For the string loop oscillation model, the mass limit $M = (7.02\pm0.22)~M_{\odot}$ introduces no restriction on the black hole spin, only the stringy parameter is restricted in this case.} The limiting values of the black hole mass are presented in Table 2. 

The mass limit given by the optical observations \citep{Bee-Pod:2002:MNRAS:} introduces an additional restriction on the GRO~J1655-40 black hole spin and on the stringy parameter $\omega$ restricting them to negative values only, as given in Table 2. We can see that due to the string loop oscillation model of the twin HF QPOs, small values of the black hole spin are forbidden, while the bottom limit of $a > 0.31$ implies possibility of fast-rotating black hole in the microquasar GRO~J1655-40, in agreement with the X-ray spectra measurements. 

}

\section{Conclusions}

We used a variety of the geodesic models of twin HF QPOs to match the twin HF QPOs simultaneously observed with a low frequency QPO in the microquasar GRO 1655-40, testing their ability to predict the GRO J1655-40 black hole mass in agreement with the limits of the optical measurements, and its spin in agreement with the spectral measurements, solving thus the recently discussed controversy \citep{Stu-Kol:2016:ASTRA:}. We have shown that the controversy in the mass and spin estimates of the GRO J1655-40 black hole can be overcome by the ER model of twin HF QPOs that can predict the black hole spin in agreement with the Fe-line spectral measurements, excluding thus the relevance of the spectral continuum measurements. However, the assumption of occurrence of the twin HF QPOs and the simultaneously observed LF QPO at a common radius has to be abandoned; the position of the source of the LF QPO relative to the position of the source of the twin HF QPOs is then determined. The ER model introduces additional restrictions on the GRO J1655-40 black hole mass ($5.4M_{\odot} < M < 5.7M_{\odot}$) and dimensionless spin ($0.94 < a < 0.96$). The radius $x_{3:2} \sim 5$ is relative distant from the black hole horizon. The radius ratio $x_{\rm nod}/x_{3:2}\sim1.6$ is then relatively low. 

Acceptable is also the special variant ER5 of the epicyclic resonant model allowing for observability of a beat frequency. In this case only the precise values of the Kerr spacetime parameters are allowed, namely, $M=5.1M_{\odot}$ and $a=0.98$. The radius $x_{3:2} \sim 3$ is closer to the black hole horizon in comparison to the case of the ER model. The radius ratio $x_{\rm nod}/x_{3:2}\sim2.6$ is then larger than in the case of the ER model. 

For comparison, we have tested also recently introduced string loop oscillation model using phenomena of non-geodesic origin that can reflect tension of toroidal magnetic fields in accretion discs \citep{Jac-Sot:2009:PHYSR4:,Kol-Stu:2010:PHYSR4:,Cre-Stu:2013:PHYSRE:}. We have demonstrated that restrictions of the string loop oscillation model on the GRO J1655-40 black hole mass can be matched to the restrictions on the mass parameter implied by the optical measurements, if the black hole spin $a>0.3$. Therefore, predictions of the string loop oscillation model can be in agreement with both the restrictions implied by the spectral measurements. The matching of the twin HF QPOs by the string loop model puts restrictions on the string-loop parameters, namely the parameter $\omega$ is restricted to negative values. The LF QPO can be matched by the relativistic nodal precession model for any black hole parameters limited by the string loop model. 

We can conclude that in order to select between the successful models of the timing effect related to the QPOs observed in the GRO J1655-40 microquasar, giving coherent limits on the black hole mass and spin, additional data from measurements of the spectral continuum and profiled spectral lines in the microquasar, and more precise data of the timing measurements of QPO, are necessary. Such precise measurements could finally exclude validity of the geodesic models for the microquasar GRO J1655-40.

We stress that confirmation or falsification of the presented QPO models can be expected due to observations of the QPOs at the microquasar GRO J1655-40 by the planned space X-ray observatory LOFT promising detection of the timing effects related to the QPOs with precision by one order higher than those obtained recently by the ROSSI X-ray detector; even the temporal evolution of QPOs during the measurements is expected in \cite{Fer-etal:2012a:ExpAstr:,Fer-etal:2012b:SPIA:}. For example, in the case of the LF QPOs with frequencies around 10Hz even lag effects could be estimated, while for the twin HF QPOs observed in hundreds of Hz the expected precision of the frequency measurements could enable to determine the character and details of the assumed resonant phenomena. 

\section*{Acknowledgments}

Z.S. acknowledges the Albert Einstein Centre for Gravitation and Astrophysics supported by the Czech Science Foundation Grant No. 14-37086G. M.K. acknowledges the Czech Science Foundation Grant No. 16-03564Y.


\section*{References}

\def\prc{Phys. Rev. C}
\def\pre{Phys. Rev. E}
\def\prd{Phys. Rev. D}
\def\jcap{Journal of Cosmology and Astroparticle Physics}
\def\apss{Astrophysics and Space Science}
\def\mnras{Monthly Notices of the Royal Astronomical Society}
\def\apj{The Astrophysical Journal}
\def\aap{Astronomy and Astrophysics}
\def\actaa{Acta Astronomica}
\def\pasj{Publications of the Astronomical Society of Japan}
\def\apjl{Astrophysical Journal Letters}
\def\pasa{Publications Astronomical Society of Australia}
\def\nat{Nature}
\def\physrep{Physics Reports}
\def\araa{Annual Review of Astronomy and Astrophysics}
\def\apjs{The Astrophysical Journal Supplement}
\def\aapr{The Astronomy and Astrophysics Review}
\def\procspie{Proceedings of the SPIE}


\end{document}